\documentclass{WileyMSP-template}
\usepackage{braket}
\usepackage{amsmath,physics,dsfont}
\usepackage{amssymb}
\usepackage{mathrsfs}
\usepackage{defs-private}

\begin{document}
\pagestyle{fancy}
\rhead{\includegraphics[width=2.5cm]{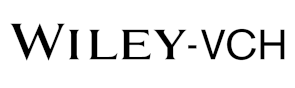}}
\title{Adaptive Variational Quantum Imaginary Time Evolution Approach for Ground State Preparation}
\maketitle

\author{Niladri Gomes}
\author{Anirban Mukherjee}
\author{Feng Zhang}
\author{Thomas Iadecola \orcid{0000-0002-5145-6441}}
\author{Cai-Zhuang Wang}
\author{Kai-Ming Ho}
\author{Peter P. Orth \orcid{0000-0003-2183-8120}}
\author{Yong-Xin Yao$^*$ \orcid{0000-0002-7830-5942}}

\dedication{}

\begin{affiliations}
N. Gomes, A. Mukherjee, F. Zhang \\
Ames Laboratory, Ames, Iowa 50011, USA

T. Iadecola, C.-Z. Wang, K.-M. Ho, P.P. Orth, Y.-X. Yao\\
Ames Laboratory, Ames, Iowa 50011, USA \\
Department of Physics and Astronomy, Iowa State University, Ames, Iowa 50011, USA \\
\textit{$^*$}ykent@iastate.edu
\end{affiliations}


\begin{abstract}
An adaptive variational quantum imaginary time evolution (AVQITE) approach is introduced that yields efficient representations of ground states for interacting Hamiltonians on near-term quantum computers. It is based on McLachlan's variational principle applied to imaginary time evolution of variational wave functions. The variational parameters evolve deterministically according to equations of motions that minimize the difference to the exact imaginary time evolution, which is quantified by the McLachlan distance. Rather than working with a fixed variational ansatz, where the McLachlan distance is constrained by the quality of the ansatz, the AVQITE method iteratively expands the ansatz along the dynamical path to keep the McLachlan distance below a chosen threshold. This ensures the state is able to follow the quantum imaginary time evolution path in the system Hilbert space rather than in a restricted variational manifold set by a predefined fixed ansatz. AVQITE is used to prepare ground states of H$_4$, H$_2$O and BeH$_2$ molecules, where it yields compact variational ans\"atze and ground state energies within chemical accuracy. Polynomial scaling of circuit depth with system size is demonstrated through a set of AVQITE calculations of quantum spin models. Finally, it is shown that quantum Lanczos calculations can also be naturally performed alongside AVQITE without additional quantum resource costs.
\end{abstract}


\section{Introduction}
Quantum computers promise to solve certain types of classically difficult problems more efficiently, with quantum simulation as an important example~\cite{nielsen2002quantum}. In the long term, given access to fault-tolerant quantum computers, adiabatic state preparation followed by quantum phase estimation may become the standard algorithm to determine the ground state energy of a quantum chemistry Hamiltonian~\cite{Aspuru05Simulated, Cao19Quantum, McArdle20Quantum}. The required circuit depth, however, is beyond capabilities of near-term noisy intermediate-scale quantum (NISQ) devices, making low-depth hybrid quantum-classical algorithm such as the variational quantum eigensolver (VQE) much more promising to achieve quantum advantage~\cite{Cao19Quantum, McArdle20Quantum, vqe, vqe_theory}. VQE takes the expectation value of a Hamiltonian, which is measured on a quantum device, as a cost function with a set of variational parameters that are optimized using classical algorithms. Excited-state calculations using VQE have also been proposed by modifying the cost function (e.g. by replacing the energy by the energy variance) or by low energy subspace expansion through linear response~\cite{vqe, mcclean2017hybrid, santagati2018witnessing, higgott2019variational, zhang2020variational, zhang2021adaptive}.

Meanwhile, quantum imaginary time evolution (QITE) has been developed as an alternative approach to prepare ground states on quantum computers~\cite{qite_chan20}. QITE inherits the advantage of classical imaginary time evolution algorithm, which allows correlations to build faster than would be allowed by the Lieb-Robinson bound that governs real time evolution~\cite{beach2019making}, and always converges to ground state~\cite{qite_chan20}. The key idea of QITE is to represent the imaginary time propagator by unitary operators via least-square fitting~\cite{qite_chan20}. The QITE circuit depth grows exponentially with the correlation domain sizes (being roughly the system's correlation length) and linearly with the number of imaginary time steps. For practical implementations, strategies to reduce circuit complexity, e.g., by utilizing symmetries and effectively combining unitaries~\cite{smqite, QITE_h2, qite_nla, sun2020quantum}, have been proposed. Meanwhile, the energy cost function of a fixed variational ansatz can be minimized following the variational QITE (VQITE) approach~\cite{theory_vqs, VQITE}, which is a special case of VQE with quantum natural gradient optimization~\cite{stokes2020quantum, koczor2019quantum}. The VQITE method has the advantage of maintaining a fixed circuit depth along the imaginary-time path, but its accuracy is limited by the fidelity of the variational ansatz in representing the ground state. While various strategies to construct variational ans\"atze have been reported during the development of VQE~\cite{hardware_efficient_vqe, vqe_theory, vqe_uccsd, kUpUCCGSD, wecker2015_trotterizedsp, qaoa}, their accuracy can be system-dependent and the variational circuits can often be suboptimal~\cite{vqe_adaptive, qcc_scott2018}. One promising strategy is to perform VQE with an adaptively generated ansatz, where operators are drawn from a predefined operator pool and iteratively added to the ansatz during the calculation~\cite{grimsleyAdaptiveVariationalAlgorithm2019}. It was shown that adaptive VQE can yield highly accurate and compact variational ans\"atze for specific problems~\cite{grimsleyAdaptiveVariationalAlgorithm2019, vqe_qubit_adaptive, yordanov2020iterative, rattew2019domain-adaptive, ostaszewski2019quantum-adaptive}. 

In this work, we develop an adaptive VQITE (AVQITE) method to perform quantum imaginary time evolution for efficient, high-fidelity ground-state preparation for interacting quantum systems. AVQITE bridges the QITE and VQITE approaches by evolving a quantum state in the system's Hilbert space towards the ground state, similar to QITE, yet with a circuit that grows sublinearly and saturates with imaginary time, leading to a final compact variational ansatz. The method generalizes the recently proposed adaptive variational quantum dynamics simulation (AVQDS) method~\cite{AVQDS} from real to imaginary time. Like AVQDS, the variational ansatz in AVQITE is automatically generated by choosing optimal multi-qubit Pauli rotation gates along the dynamical path to keep a measure of ansatz quality, the McLachlan distance~\cite{theory_vqs, variational_mclachlan}, within a desired accuracy. We demonstrate the capabilities of AVQITE calculations by preparing the ground states of an H$_4$ chain and H$_2$O and BeH$_2$ molecules at representative bond lengths with increasing electron correlation effects. We find total energies to chemical accuracy with compact ans\"atze similar to qubit-ADAPT-VQE results~\cite{vqe_qubit_adaptive}, yet without resorting to the complex optimization of parameters in a high-dimensional nonconvex energy landscape. Furthermore, Quantum Lanczos eigenvalue calculations can be carried out together with AVQITE. We additionally demonstrate the favorable polynomial system-size scaling of AVQITE calculations for local spin models. We envision AVQITE, with its compact variational circuits and avoidance of explicit high-dimensional optimization, as a viable way to efficiently prepare ground states of interacting fermion systems (e.g., molecules) on NISQ devices.

\section{AVQITE Algorithm}
The AVQITE algorithm generalizes the recently introduced AVQDS approach~\cite{AVQDS} from real to imaginary time evolution. By time evolving a quantum state in imaginary time, it yields the ground state of a system as the final state. While the derivation of AVQITE resembles that of AVQDS, there are a few key differences that we point out in the following. 

\subsection{Variational Quantum Imaginary Time Evolution method}
\subsubsection{Algorithm}
The theory of VQITE has been developed in reference~\cite{theory_vqs} and has been used to find ground-state energies of H$_2$ and LiH molecules on classical simulators~\cite{theory_vqs, VQITE}. Here, we review the VQITE formalism within the density matrix approach, which is insensitive to the global phase of the quantum state. Consider a system with Hamiltonian $\h$ in a pure state $\ket{\Psi}$. The evolution of the density matrix $\hat{\rho}\equiv \dyad{\Psi}$ under the imaginary-time propagator $e^{-\tau\h}$ is governed by the Liouville–von Neumann-type equation~\cite{theory_vqs, berman1991Liouville-vonNeumann}
\be
\frac{d\hat{\rho}}{d \tau} = \Lag[\hat{\rho}],
\ee
where the superoperator $\Lag[\hat{\rho}] = -\{\h, \hat{\rho} \} + 2\av{\h} \hat{\rho}$ with the anticommutator $\{\h, \hat{\rho} \} = \h\hat{\rho} + \hat{\rho}\h$, and the Hamiltonian expectation value $\av{\h} = \Tr[\hat{\rho} \h]$. The so-called imaginary time $\tau \in \mathbb{R}$ is a real positive parameter. For a generic variational ansatz $\ket{\Psi[\bth]}$ with a real parameter vector $\bth$ of dimension $N_{\bth}$, the squared McLachlan distance $L^2$ is defined by the Frobenius norm of the difference between the variational and exact state propagations along the imaginary time axis, i.e.~\cite{theory_vqs}:
\bea
L^2 &\equiv&\norm{\sum_{\mu=1}^{N_\theta} \frac{\partial \hat{\rho}[\bth]}{\partial \theta_\mu} \dot{\theta}_\mu - \Lag[\hat{\rho}]}^2 \notag \\
&=&\sum_{\mu \nu} M_{\mu \nu} \dot{\theta}_\mu \dot{\theta}_\nu -2 \sum_\mu V_\mu \dot{\theta}_\mu + \Tr[\Lag[\hat{\rho}]^2], \label{L2}
\eea
which is a quadratic function of the time derivatives $\{\dot{\theta}_\mu \equiv  \partial \theta_\mu / \partial \tau \}$.
The real symmetric $N_{\bth} \times N_{\bth}$ matrix $M$ is specified as
\be
M_{\mu \nu} = 2 \Re\left[\frac{\partial \bra{\Psi[\bth]}}{\partial \theta_\mu} \frac{\partial \ket{\Psi[\bth]}}{\partial \theta_\nu} + \frac{\partial \bra{\Psi[\bth]}}{\partial \theta_\mu}\ket{\Psi[\bth]} \frac{\partial \bra{\Psi[\bth]}}{\partial \theta_\nu}\ket{\Psi[\bth]}\right], \label{eq: M}
\ee
which is equivalent to the quantum Fisher information matrix~\cite{stokes2020quantum, koczor2019quantum, meyer2021fisher}. The real vector $V$ of dimension $N_{\bth}$ is defined as
\bea
V_\mu &=& 2\Re\left[-\frac{\partial \bra{\Psi[\bth]}}{\partial \theta_\mu}\h\ket{\Psi[\bth]} + \bra{\Psi[\bth]}\frac{\partial \ket{\Psi[\bth]}}{\partial \theta_\mu}\av{\h}_{\bth} \right] \notag \\
&=& 2\Re\left[-\frac{\partial \bra{\Psi[\bth]}}{\partial \theta_\mu}\h\ket{\Psi[\bth]}\right] \label{eq: V}
\eea
with the abbreviation $\av{\h}_{\bth}\equiv \Av{\Psi[\bth]}{\h}$. The second term in the first line of equation~\eqref{eq: V} vanishes due to the normalization $\olp{\Psi[\bth]}=1$. Therefore, $V$ is equal to the energy gradient vector. The final term in Eq.~\eqref{L2} can be expressed in terms of the energy variance as
\bea
\Tr[\Lag[\hat{\rho}]^2] &=& 2\left(\av{\h^2}_{\bth} - \av{\h}_{\bth}^2 \right) = 2 \,  \text{var}_{\bth}[\h]. \label{eq: var}
\eea

The McLachlan variational principle amounts to minimizing the quadratic cost function $L^2$ with respect to $\{\dot{\theta}_\mu \}$. This leads to the following linear equations of motion:
\be
\sum_\nu M_{\mu \nu}\dot{\theta}_\nu = V_\mu, \label{eq: eom}
\ee
which has the same form as the parameter update rule arising within the quantum natural gradient approach~\cite{stokes2020quantum, koczor2019quantum}. The second term in equation~\eqref{eq: M} originates from the global phase of the wavefunction~\cite{VQITE, theory_vqs}. Below we adopt a purely real wave function ansatz in a pseudo-Trotter form (with odd number of Pauli-Y terms) for which the global phase contribution vanishes. Consequently, the density-matrix and wavefunction-based derivations reach exactly the same results. 

The optimal McLachlan distance $L^2$ of the variational ansatz $\Psi[\bth]$ given by
\be
L^2 = 2\, \text{var}_{\bth}[\h]-\sum_{\mu \nu}V_\mu M^{-1}_{\mu \nu} V_\nu. \label{eq: L2min}
\ee
VQITE is formulated in the exactly same form as real time variational quantum dynamics simulations (VQDS)~\cite{theory_vqs, AVQDS}, with exception of the definition of vector $V$ in Equation~\eqref{eq: V} due to the different superoperator $\Lag[\hat{\rho}]$ in the Liouville–von Neumann-type equation. 

\begin{figure*}[t]
	\centering
	\includegraphics[width=\linewidth]{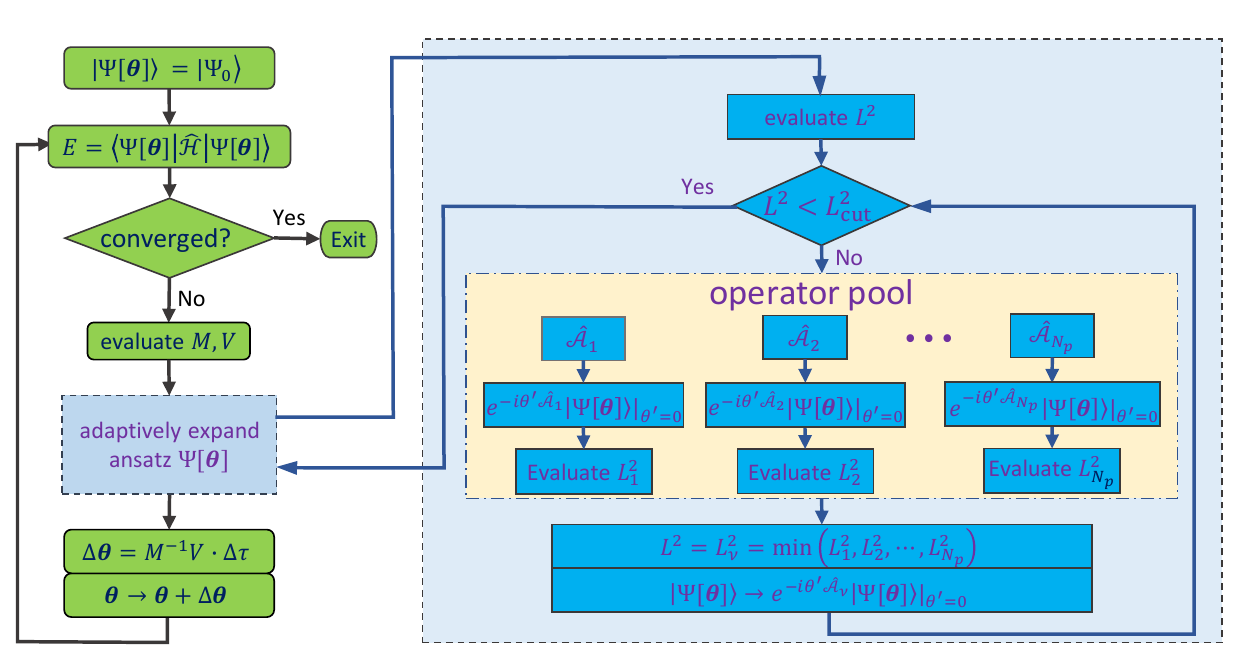}
	\caption{
	\textbf{Schematic illustration of variational quantum imaginary time evolution algorithm, with an additional module to adaptively expand the ansatz.} The green flowchart on the left shows a typical VQITE calculation. In AVQITE, a module (blue) is introduced to adaptively expand the variational ansatz by selectively appending parametric rotation gates to keep the McLachlan distance $L^2$ under a threshold $L^2_\text{cut}$ along the imaginary-time evolution path.
	}
	\label{fig: vqite}
\end{figure*}

\subsubsection{Flowchart}
A typical VQITE calculation, which integrates the equation of motion~\eqref{eq: eom} with a constant time step $\Delta\tau$ according to the Euler method, is illustrated in the green charts of Figure~\ref{fig: vqite}. Given a fixed variational ansatz initialized to a reference state $\ket{\Psi_0}$ that is easily prepared on a quantum computer, the energy expectation value is first measured. The convergence criterion, such as the energy difference between two consecutive steps, is checked. If the convergence condition is not satisfied, the real symmetric matrix $M$ in Equation~\eqref{eq: M} and vector $V$ in Equation~\eqref{eq: V} are determined, with which the step size $\Delta\bth$ of the variational parameter vector $\bth$ at time step $\Delta\tau$ is calculated. The ansatz state $\ket{\Psi[\bth]}$ with the updated parameters triggers another iteration until convergence is reached. Note that the energy variance var$_{\bth}[\h]$ provides a quality measure of the variational ansatz in approximating the ground state.

\subsection{Adaptive Variational Quantum Imaginary Time Evolution Method}
\subsubsection{Algorithm and Flowchart}
The VQITE formalism is presented for a generic wave function ansatz. In the development of VQE, variational ans\"atze $\ket{\Psi[\bth]} = \hat{\U}[\bth]\ket{\Psi_0}$ with two different forms of the unitary operator $\U[\bth]$ acting on a reference state $\ket{\Psi_0}$ have been proposed. In hardware efficient ans\"atze, the unitary operator $\hat{\U}[\bth]$  is a product of parametrized native gates of the real device, e.g., single-qubit rotation gates plus two-qubit entangling gates~\cite{hardware_efficient_vqe}. Alternatively, at a higher algorithmic level, $\hat{\U}[\bth]$ can be expressed as a product of $N_{\bth}$ multi-qubit rotation gates in a pseudo-Trotter form:
\be
\ket{\Psi[\bth]} = \prod_{\mu=1}^{N_{\bth}} e^{-i\theta_\mu \hat{\A}_\mu} \ket{\Psi_0}, \label{eq: ansatz}
\ee
where $\hat{\A}_\mu$ are Hermitian operators. The unitary coupled cluster ansatz and its variants~\cite{vqe_theory, vqe_uccsd, kUpUCCGSD}, the Hamiltonian variational ansatz~\cite{wecker2015_trotterizedsp}, and the ansatz in the quantum approximate optimization algorithm (QAOA) all belong to this category~\cite{qaoa}. In AVQITE, we adopt a variational ansatz in the above form~\eqref{eq: ansatz}, and allow the number of $\hat{\A}$-operators to be dynamically expanded along the imaginary-time evolution path to maintain high accuracy in representing the evolving quantum state, as shown in the blue module of Figure~\ref{fig: vqite}. 

The initial steps of an AVQITE calculation are the same as those of a VQITE calculation, with the exception of great flexibility in the choice of initial ansatz $\ket{\Psi[\bth]}$. While the simulation accuracy is tied to the initial ansatz in VQITE calculations, an AVQITE calculation monitors the quality of $\ket{\Psi[\bth]}$ in representing the evolving quantum state and adaptively expands the form of $\ket{\Psi[\bth]}$ to maintain a fixed accuracy. In fact, AVQITE calculations can simply take any $\ket{\Psi_0}$ as the initial ansatz. For convenience, we use a product state as our initial state in all our calculations below, which is easily prepared on a quantum processor unit (QPU). In the first iteration where no variational parameters are present, the McLachlan distance $L^2$ in Equation~\eqref{eq: L2min} is determined by the energy variance var$[\h]$ in state $\ket{\Psi_0}$, which is generally larger than the threshold $L^2_\text{cut}$, since $\ket{\Psi_0}$ is not an eigenstate of $\h$. As a result, a predefined operator pool is scanned and an operator $\hat{\A}_\nu$ is chosen to construct a unitary $e^{-i\theta' \hat{\A}_\nu}$ to be appended to the variational ansatz $\ket{\Psi[\bth]}$, which produces minimal McLachlan distance, as illustrated in Figure~\ref{fig: vqite}. The additional parameter $\theta'$ associated with the new operator is always initialized to zero to keep the imaginary time evolution of the variational state continuous. Nevertheless, the McLachlan distance can still be reduced by addition of $e^{-i \theta' \hat{\A}_\nu}$, because it involves derivatives of the ansatz. The adaptive procedure of selectively appending a new unitary to the ansatz continues until the updated McLachlan distance satisfies $L^2 < L^2_\text{cut}$. The expanded variational parameter vector $\bth$ is subsequently updated at the imaginary time step as in VQITE, and new iterations proceed until energy convergence is reached. 

\subsubsection{Important Technical Details}
The accuracy of AVQITE calculations is controlled by the McLachlan distance threshold $L^2_\text{cut}$, while the ability to reach a compact final ansatz $\ket{\Psi[\bth]}$ is tied to the operator pool. For quantum chemistry calculations, different ways to efficiently construct operator pools and pool completeness conditions have been extensively discussed in the context of adaptive approaches to VQE~\cite{grimsleyAdaptiveVariationalAlgorithm2019, vqe_qubit_adaptive, yordanov2020iterative}. The fermionic operator pool proposed in  reference~\cite{grimsleyAdaptiveVariationalAlgorithm2019} is composed of the single excitation operators and double excitation operators with respect to a Hartree-Fock (HF) reference state. When translated to a qubit representation using encoding methods such as Jordan-Wigner (JW) mapping~\cite{map_jw}, a single excitation operator can result in a weighted sum of two Pauli strings, and six Pauli strings for a double excitation. Here a Pauli string is defined as a product of single qubit Pauli operators, namely $X, Y$ and $Z$. Alternatively, a qubit operator pool can also be constructed directly with rudimentary Pauli strings~\cite{vqe_qubit_adaptive, yordanov2020iterative}. Compared with fermionic operator pools, ADAPT-VQE calculations with qubit operator pools generate variational ans\"atze with considerably shallower circuits at the price of more variational parameters. As the explicit complex nonconvex optimization problem in the high-dimensional parameter space of VQE is avoided in AVQITE, the qubit operator pool is therefore very appealing and adopted in the following AVQITE calculations. In the AVQITE calculations, we construct a qubit operator pool by choosing all the Pauli strings present in the fermionic single and double excitation operators of the unitary coupled cluster (UCCSD) ansatz~\cite{vqe_theory, vqe_uccsd}. The parity mapping is used to transform the fermionic excitation operators to qubit operators~\cite{map_bk, seeley2012bravyi}, as fermion to qubit mappings other than JW have not been studied before in the context of operator pool construction~\cite{vqe_qubit_adaptive}. Since the fermionic excitation operators are real, every Pauli string in the operator pool contains an odd number of Pauli $Y$ operators and is therefore antisymmetric. The unitaries in equation~\eqref{eq: ansatz} are thus real and an initially real wave function (i.e. with real coefficients) remains real when evolving along the imaginary time path. Consequently, the expression $\frac{\partial \bra{\Psi[\bth]}}{\partial \theta_\mu} \ket{\Psi[\bth]} = \bra{\Psi_{\mu-1}}i \hat{\mathcal{A}}_\mu\ket{\Psi_{\mu-1}}$ in Eq.~\eqref{eq: M}, where $\ket{\Psi_\mu[\theta]} = \prod_{\mu' = 0}^\mu e^{-i \theta_{\mu'} \hat{\mathcal{A}}_{\mu'}} \ket{\Psi_0}$, vanishes for any $\mu$. Therefore, the second term of $M$ in Eq.~\eqref{eq: M} which originates from the global phase of the wavefunction vanishes, and the density-matrix and wavefunction-based approaches lead to exactly the same results.

To evolve the quantum state $\ket{\Psi[\bth]}$ along the imaginary time path by updating parameters as 
\be 
\bth \to \bth + \dot{\bth} \Delta\tau = \bth + M^{-1}V \Delta\tau,
\ee
the imaginary time step $\Delta\tau$ needs to be small enough to maintain high numerical accuracy, while being sufficiently large for fast convergence. In practical AVQITE calculations of molecules (with energy measured in Hartree atomic units) and local spin models (with energy measured in units of the spin-spin interaction $J$), we find $\Delta\tau = 0.1$ works well, leading to a ground-state solution with relatively fewer steps. To stabilize the inversion of the matrix $M$ against potential proximity to singularity in numerical simulations, we adopt the Tikhonov regularization approach by adding a small diagonal element $\xi = 10^{-6}$ to $M$. Alternatively, the matrix inversion problem can be avoided by solving the linear equation~\eqref{eq: eom} using optimization techniques~\cite{endo2020calculation}.

\subsubsection{Implementation Strategies and Measurement Costs on Real Devices}
\label{sec: implementation}
The implementation of AVQITE amounts to measuring the symmetric matrix $M$ in Eq.~\eqref{eq: M}, the vector $V$ in Eq.~\eqref{eq: V} and the scalars $\av{\h}_{\bth}$, $\av{\h^2}_{\bth}$ on a quantum device, all of which have been discussed in the context of VQITE in reference~\cite{vqe_theory, VDynamics_Li}. Specifically, the energy $\av{\h}_{\bth}$ can be obtained as a weighted sum of expectation values of the Pauli strings in the Hamiltonian, a procedure often termed ``Hamiltonian averaging"~\cite{vqe_theory}:
\be
\av{\h}_{\bth} = \sum_i h_i \av{\hat{O}_i}_{\bth}. 
\ee
Here $h_i$ and $\hat{O}_i$ are the coefficient and Pauli string of the $i$th Hamiltonian term in the qubit representation. The expectation value $\av{\h^2}_{\bth}$ can be directly measured similarly by replacing $\h$ with $\h^2$. The circuit implementation to measure $V$, which is essentially the energy gradient, has also been discussed in the context of VQE optimization~\cite{alan_ucc2018}. The proposed indirect measurement circuit introduces an ancillary qubit with a Hadamard-type test. Two controlled-unitary gates, specifically, controlled multi-qubit Pauli gates, need to be implemented. Alternatively, general strategies to replace indirect measurements by direct measurements have also been proposed~\cite{mitarai2019methodology}. It is especially appealing to use the parameter-shift rule to evaluate the energy gradient $V$~\cite{mari2021estimating, li2017hybrid, mitarai2018quantum, schuld2019evaluating}, as it only requires some additional measurements of Hamiltonian expectation values at shifted parameters, without the necessity of introducing new circuits. For the matrix element $M_{\mu\nu}=2\Re\left[\frac{\partial \bra{\Psi[\bth]}}{\partial \theta_\mu} \frac{\partial \ket{\Psi[\bth]}}{\partial \theta_\nu}\right]$ in Eq.~\eqref{eq: M}, where we have used that the global phase term vanishes with the pseudo-Trotter ansatz~\eqref{eq: ansatz}, the diagonal elements can be simplified to $M_{\mu\mu}=2\Av{\Psi_{\mu-1}[\bth]}{\hat{\A}_{\mu-1}^2}$, where  $\ket{\Psi_{\mu}[\bth]}=\hat{\U}_{0,\mu}[\bth]\ket{\Psi_0}$ with $\hat{\U}_{j,k}[\bth] \equiv \prod_{\mu'=j}^{k} e^{-i\theta_{\mu'} \hat{\A}_{\mu'}}$. This can be measured with the Hamiltonian averaging method for the Hermitian operator $\hat{\A}_{\mu}^2$. Because $\hat{\A}_{\mu}$ is a single Pauli string for the qubit operator pools adopted here, $\hat{\A}_{\mu}^2$ is an identity operator and $M_{\mu\mu}=2$. The off-diagonal elements of $M$ can be simplified to $M_{\mu\nu}=2\Re\left[\bra{\Psi_{\mu}[\bth]} \hat{\A}_{\mu} \hat{\U}_{\mu+1,\nu-1}[\bth] \hat{\A_{\nu}}\ket{\Psi_{\nu-1}[\bth]}\right]$ for $\mu < \nu$, which can be measured using generalized Hadamard test circuit~\cite{theory_vqs, VDynamics_Li}, or direct measurement circuits~\cite{mitarai2019methodology}.

Generally for an $N$-qubit system with Hamiltonian $\h$ composed of $N_\text{H}$ Pauli strings, and a parameterized ansatz $\ket{\Psi[\bth]}$ of $N_{\bth}$ parameters with an operator pool of dimension $N_\text{p}$, the upper bound for the number of distinct measurement circuits $N_\text{m}$ for AVQITE calculations is given by $N_\text{H} + N_\text{H}^2 + N_{\bth}(N_{\bth}-1)/2 + N_\text{p}N_{\bth}$, where $N_\text{p}N_{\bth}$ comes from the operator selection step. It consists of direct measurements of $\h$, $\h^2$, and Hadamard tests to measure matrix $M$. Here we assume the gradient vector $V$ is evaluated using the parameter shift rule~\cite{mari2021estimating, li2017hybrid, mitarai2018quantum, schuld2019evaluating}. Assuming a polynomial scaling of $N_\text{H} \propto N^h$, $N_\text{p} \propto N^p$, and $N_{\bth} \propto N^q$, the leading order for distinct measurement circuits becomes $N_\text{m} \propto N^{max(2h,2q,p+q)}$. For the local spin chain model calculations with local operator pool (see Sec.~\ref{sec:SpinModels}), both $N_\text{H}$ and $N_\text{p}$ scale linearly with system size $N$. Therefore, the number of measurement circuits scales as $N_\text{m} \propto N^{\max(2,1+q)}$. Quantum chemistry calculations (see Sec.~\ref{sec:Chemistry}) are usually more challenging, with $N_\text{H} \propto N^4$ and $N_\text{p} \propto N^4$ for the adopted UCCSD pool, which leads to $N_\text{m} \propto N^{max(8,2q,q+4)}$. The $N^8$ scaling for direct measurement of $\h^2$ and the $N^{q+4}$ scaling for operator screening can be limiting factors. Remarkably, linear-scaling ($p=1$) operator pools have also been proposed in the context of quantum chemistry calculations~\cite{vqe_qubit_adaptive}. (Nevertheless, we leave the study of AVQITE calculations with different pools for future work.) Furthermore, the direct measurement circuits for $\av{\h^2}_{\bth}$ can be replaced with the quantum power method for more NISQ-friendly scaling~\cite{seki2021quantumpower}, where the $\h^2$ is approximated as a linear combination of unitaries.  Alternatively, the evaluation of $\av{\h^2}_{\bth}$ can be completely avoided by modifying the McLachlan distance criterion. We observe that while it is necessary to have $\av{\h^2}_{\bth}$ to evaluate the McLachlan distance $L^2$, the estimate of the decrease of $L^2$ upon appending a new unitary to the ansatz is independent of $\av{\h^2}_{\bth}$. Therefore, the criterion $L^2 < L^2_\text{cut}$ can be replaced with a $\av{\h^2}_{\bth}$-free criterion based on the maximal reduction of $L^2$ due to an added unitary generated by an operator in the operator pool, i.e.,
\be
\max_{\hat{\A}_\nu \in \mathscr{P}} (L^2-L^2[\hat{\A}_\nu]) < \Delta_{L^2}^\text{cut}\,. 
\ee
However, this new criterion implies that screening through all operators in the pool $\mathscr{P}$ becomes necessary at every time step. 

By adopting the above way to circumvent the measurement of $\av{\h^2}_{\bth}$, the main measurement cost for AVQITE calculation at each time step is tied to measuring matrix $M$ and gradient $V$. The measurement cost analysis in this case closely follows Ref.~\cite{van2020measurement} for generic metric-aware variational quantum algorithms. For simplicity, we consider a uniform distribution of measurements among the $N_{\bth}^2+N_{\bth}N_\text{p}$ elements of $M$ and $N_{\bth}+ N_\text{p}$ elements of $V$, where the $N_\text{p}$-dependent parts come from the operator screening step in the adaptive ansatz expansion procedure. To reduce the uncertainty in evaluating $\dot{\bth} = M^{-1}V$ due to quantum noise to a precision $\epsilon$, i.e., $\sum_{i=1}^{N_{\bth}}\text{Var}[\dot{\theta}_i] = \epsilon^2$, the total number $\mathcal M$ of measurements is upper bounded as:
\bea
\M &\le& 2\epsilon^{-2}\text{Spc}[M^{-1}]^2|V|_\text{max}^2f_\text{F}N_{\bth}^4(1+\frac{N_\text{p}}{N_{\bth}}) \notag \\
&+& 2\epsilon^{-2}\text{Spc}[M^{-1}]\text{Spc}[\h]f_\text{V}N_{\bth}^2(1+\frac{N_\text{p}}{N_{\bth}}).
\eea
Compared with the measurement cost for a fixed ansatz (see Eqs. (3) and (4) in Ref.~\cite{van2020measurement}), a factor of $(1+\frac{N_\text{p}}{N_{\bth}})$ is introduced due to the additional $N_{\bth}N_\text{p}$ measurements for $M$ and $N_\text{p}$ measurements for $V$ in the operator selection step. The first term in the above equation is associated with measurements of $M$, and the second with measurements of $V$. $\text{Spc}[\h]=\sum_{i}h_i^2$ is the Hilbert-Schmidt scalar product of $\h$. $\text{Spc}[M^{-1}]$ denotes the average of the squared singular values of $M^{-1}$, which is upper bounded by the Tikhonov regularization parameter $\xi$ as $\text{Spc}[M^{-1}]\le \xi^{-2}$. $|V|_\text{max}$ is the largest absolute value of all elements of $V$. Because the generator of a multi-qubit Pauli rotation gate is a single Pauli string in the current calculations, the constant factor $f_\text{F}$ obeys $f_\text{F} \le 2$~\cite{van2020measurement}. The constant factor $f_\text{V}$ depends on the Hamiltonian measurement circuits, and is bounded as $1\le f_\text{V}\le N_\text{H}$. In the above analysis, the indirect Hadamard test circuits are used to measure the gradient $V$. The measurement cost for estimating the quantum Fisher information matrix $M$ relative to that for $V$ is shown to be asymptotically negligible with increasing number of iterations, due to the diminishing gradient $|V|_\text{max}$~\cite{van2020measurement}.



\subsubsection{Quantum Lanczos Calculation}
The focus of the QITE approach lies on the final quantum state, which converges to the ground state for large enough time. Within QITE, all previous quantum states along the path are discarded. In contrast, the quantum Lanczos (QL) method was developed to achieve a more efficient calculations of ground and excited state energies by exploiting information contained in all quantum states along the path~\cite{qite_chan20}. The essence of QL is to diagonalize the Hamiltonian within the Krylov subspace spanned by a subset of imaginary time states $\{ \ket{\Psi_{n}} = C_n e^{-n\Delta\tau\h}\ket{\Psi_0}\}$ with even time step indices $n=0,2,\dots N$ and normalization constants $C_n$. In the classical Lanczos algorithm, the reduced Hamiltonian in the Krylov subspace is brought to a tridiagonal form by sequentially applying the Hamiltonian on orthonormalized Krylov basis vectors~\cite{demmel1997applied}. In contrast, QL directly represents the reduced Hamiltonian $\h$ in the normalized basis $\{ \ket{\Psi_{n}}\}$ characterized by an overlap matrix $S_{n n'} =\frac{C_n C_{n'}}{C^2_{(n+n')/2}}$. As the normalization coefficient $C_{(n+n')/2}$ is used, only the even-$n$ imaginary-time states (for which $(n+n')/2$ is an integer) are chosen for the construction of the Krylov subspace. Accordingly, the dense Hamiltonian matrix elements can be evaluated as $H_{n n'} = S_{n n'} E_{(n+n')/2}$, where the expectation value $E_{n} \equiv H_{n n} = \Av{\Psi_n}{\h}$. The normalization coefficient $C_n$ can be calculated recursively as $C_n^{-2} = C_{n-1}^{-2}(1-2\Delta\tau E_{n-1} + \bigO(\Delta \tau^2))$ with $C_0=1$. Therefore, only the expectation values of the Hamiltonian $\{E_n \}$ are needed to set up the generalized eigenvalue equation in the Krylov subspace for QL. As $\{E_n \}$ are readily available in an AVQITE calculation, a QL calculation can be efficiently performed alongside AVQITE to get the energy eigenvalues with no additional quantum resource costs. Nevertheless, as the Lanczos eigenvector is expressed as a linear combination of imaginary time states, it can be much more involved to measure expectation values of other observables beyond energy on a quantum computer~\cite{childs2012hamiltonian}.

\begin{figure*}[t]
	\centering
	\includegraphics[width=\linewidth]{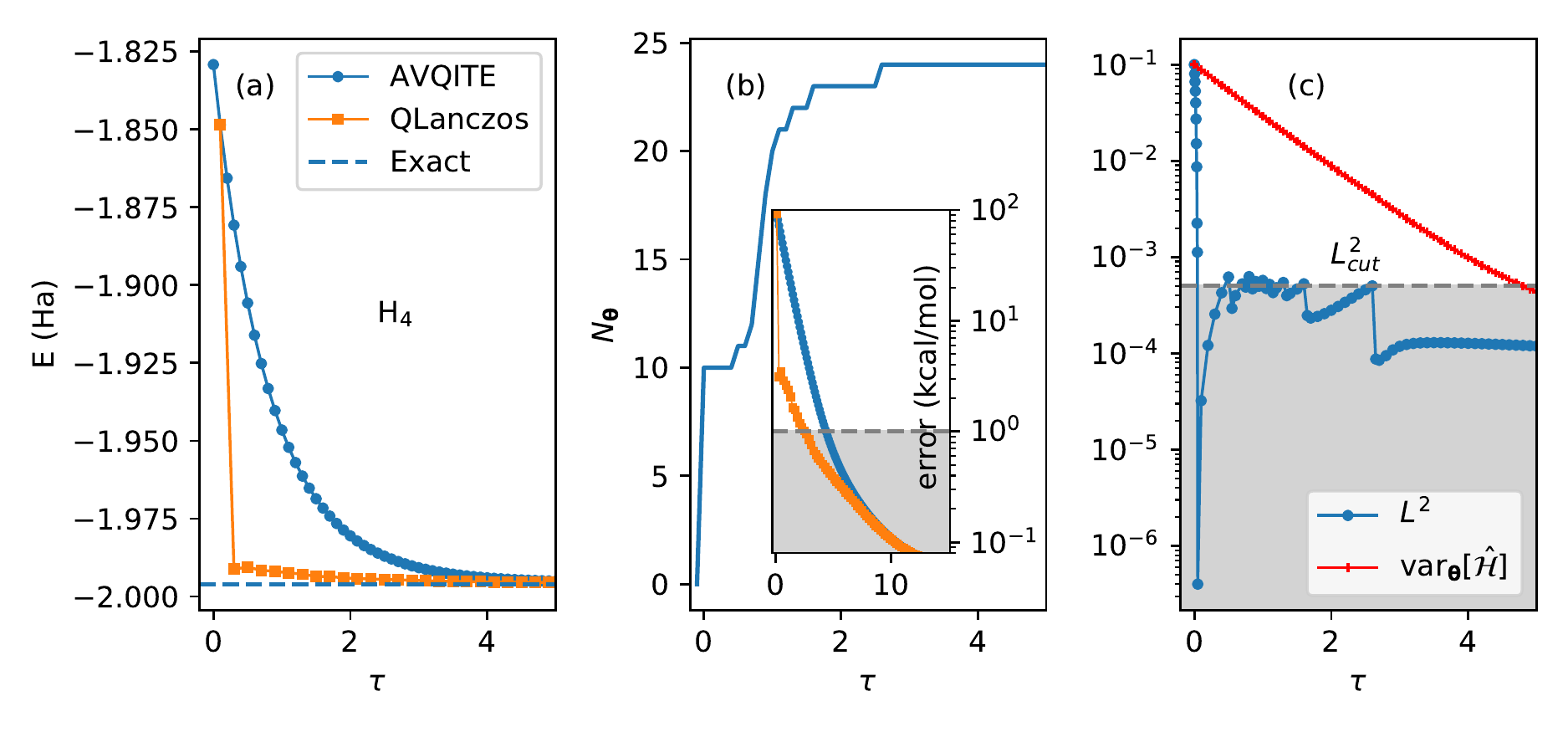}
	\caption{
	\textbf{AVQITE calculation for H$_4$ chain.} Along the imaginary time path of $\tau=N\Delta\tau$, the total energies of H$_4$ from AVQITE and quantum Lanczos calculations are shown in panel (a), the number of variational parameters $N_{\bth}$ in (b), and the McLachlan distance $L^2$ and energy variance var$_{\bth}[\h]$ in (c).
	The exact full configuration interaction result is shown as the dashed line in (a) for reference. The total energy errors $E-E_\text{Exact}$ are plotted in the inset of (b), with the shaded area denoting chemical accuracy. The chosen threshold $L^2_\text{cut}=5\times 10^{-4}$ is indicated by a dashed line in (c).
	}
	\label{fig: h4}
\end{figure*}

\section{AVQITE calculations of molecules}
\label{sec:Chemistry}

The \textit{ab initio} nonrelativistic molecular electron Hamiltonian is given by 
\be
\h = \sum_{p q}\sum_{\sigma}h_{p q}\cc_{p\sigma}\ca_{q\sigma} + \frac{1}{2}\sum_{p q r s}\sum_{\sigma \sigma'}h_{p q r s}\cc_{p \sigma} \cc_{r \sigma'} \ca_{s \sigma'} \ca_{q \sigma}, \label{eq: mh}
\ee
where the one-electron core part $h_{p q} = \int d\br \phi_{p}^{*}(\br)(\T + \V_{ion})\phi_q(\br)$, and the two-electron Coulomb integral is obtained as
\be
h_{p q r s} = \iint d\br d\br' \phi_{p}^{*}(\br)\phi_{r}^{*}(\br')\V_{e e}\left( \abs{\br - \br'}\right)\phi_s(\br')\phi_q(\br).
\ee
Here $p,q,r,s$ are composite indices for atom and orbital, and $\sigma$ is the spin index. $\T$, $\V_{ion}$ and $\V_{e e}$ are the kinetic energy, ionic potential operator and  Coulomb interaction operator, respectively. The standard STO-3G minimal basis set is adopted for the basis orbital functions $\{\phi(\br)\}$. In the following AVQITE calculations, the \textit{PySCF} quantum chemistry package is first used to generate the molecular Hamiltonian~\eqref{eq: mh} and produce the restricted Hartree-Fock(HF) solution~\cite{pyscf_sun2018}. A basis transformation from atomic orbitals to molecular orbitals is performed to the Hamiltonian~\eqref{eq: mh} for the convenience of preparation of the initial HF state as a tensor product state on the quantum computer. The parity transformation is applied to get the qubit representation of the molecular Hamiltonian, where two qubits are tapered due to the $Z_2$ symmetry from total electron number and total spin conservation~\cite{map_bk, map_three, bravyi2017tapering}. All the molecular calculations reported below, including AVQITE and qubit-ADAPT-VQE, are performed in this representation.

\begin{figure*}[t!]
	\centering
	\includegraphics[width=\linewidth]{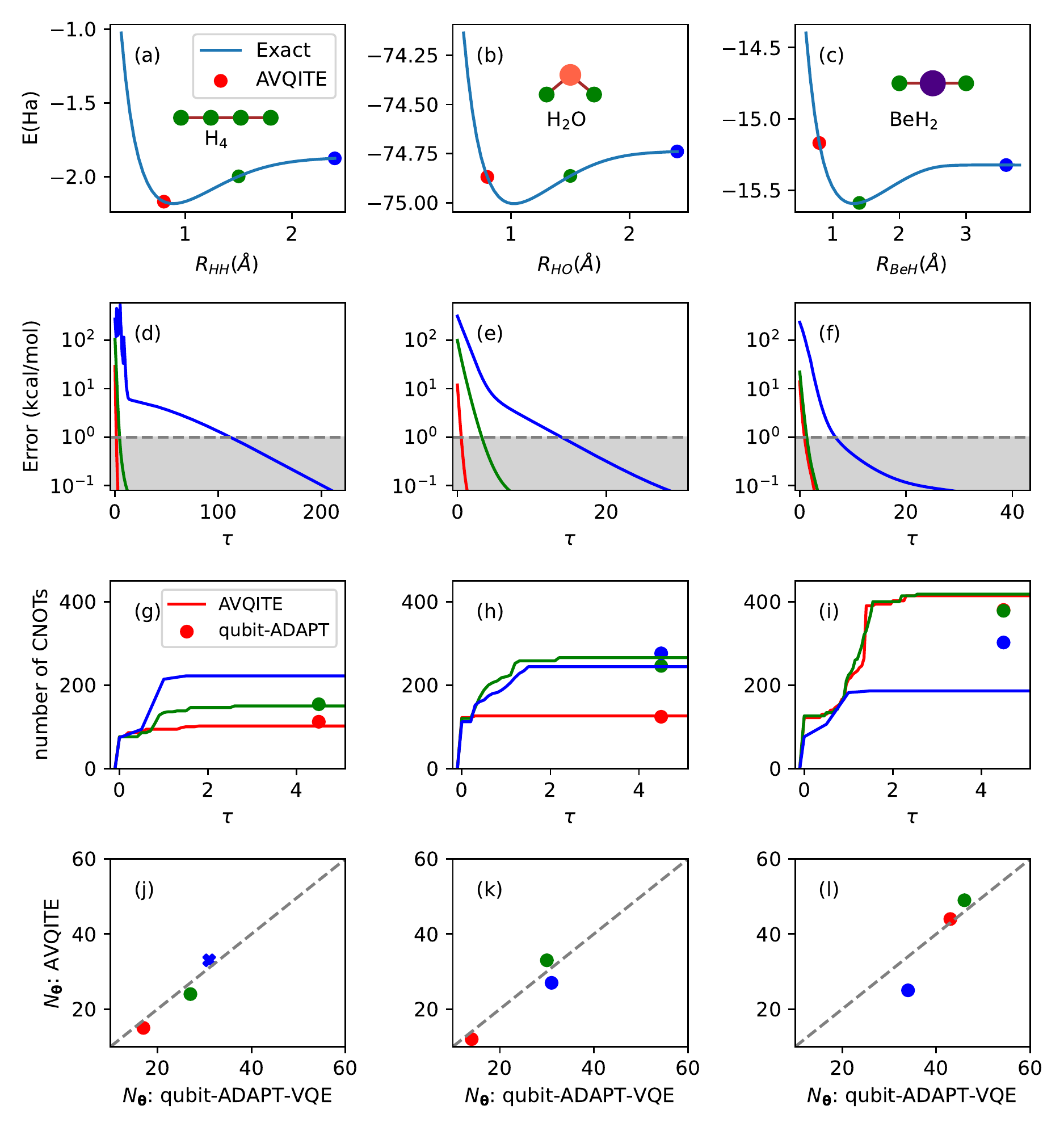}
	\caption{
	\textbf{AVQITE calculations of molecules with varying electron correlations.} (a-c): AVQITE total energies of H$_4$, H$_2$O and BeH$_2$ molecules at three bond lengths colored by red, green and blue. The exact potential energy curves from FCI calculations are also shown for reference. (d-f): AVQITE energy error, $E-E_{Exact}$, as a function of imaginary time step $\tau=N\Delta\tau$ for the molecules at bond lengths with the same color coding. The shaded area indicates errors smaller than chemical accuracy. (g-i): solid line shows the dependence of number of CNOT gates in the AVQITE state preparation circuit as a function of imaginary time for each set of molecules, with solid circle indicating the number of CNOT gates of the final converged qubit-ADAPT-VQE ans\"atze. (j-l): Number of variational parameters $N_{\bth}$ of the AVQITE ansatz vs that for qubit-ADAPT-VQE when the energy accuracy reaches 0.1 mHa. The symbol ``$\times$'' indicates the qubit-ADAPT-VQE calculation does not reach chemical accuracy. The dashed line is a reference for equal number of parameters.
	}
	\label{fig: all}
\end{figure*}

In Figure~\ref{fig: h4} we illustrate a detailed numerical AVQITE calculation of an H$_4$ chain molecule at a uniform bond length $R=1.5$\AA, where significant electron correlation effects are present. Starting with the initial Hartree-Fock state $\ket{\Psi_0}$, the total energy monotonically decreases and converges toward the exact result with increasing imaginary time $\tau=N\Delta\tau$ at $\Delta\tau=0.1$, as shown by the blue circles in Figure~\ref{fig: h4}(a). The associated error, which is defined as the difference between the AVQITE energy and the exact result from a full configuration interaction (FCI) calculation $E-E_\text{Exact}$, is shown in the inset of Figure~\ref{fig: h4}(b) on a log scale. The AVQITE calculation reaches chemical accuracy of 1 kcal/mol after $\tau = 4.7$. The quantum state fidelity~\cite{nielsen2002quantum} $f \equiv \abs{\ov{\Psi[\bth]}{\Psi_\text{Exact}}}^2 $, i.e., the squared overlap between the ansatz state and the exact ground state, goes beyond $99.9\%$. The number of variational parameters $N_{\bth}$, or equivalently the number of multi-qubit Pauli rotation gates, increases at several imaginary time steps and flattens at 24. For comparison, the qubit-ADAPT-VQE calculation of H$_4$ using the same operator pool generates a final ansatz of 30 variational parameters upon reaching chemical accuracy. Specific information about the McLachlan distance $L^2$ is plotted in Figure~\ref{fig: h4}(c), where $L^2$ is reduced below the threshold $L^2_\text{cut}=5\times 10^{-4}$ by adaptively expanding the variational ansatz whenever the initial value $L^2 > L^2_\text{cut}$ at any time step. Together with $L^2$, the energy variance var$_{\bth}[\h]$ along the imaginary path is shown to be almost linear on a semi-log scale, implying an exponential convergence. In fact, the AVQITE total energies $E(\tau)$ can also be fitted with an exponential function, $E(\tau) = E_{\infty} + e^{-a\tau}$, with $E_\infty =-1.9957$ Ha in the infinite time step limit, which is within chemical accuracy with an error of 0.2 kcal/mol. Alternatively, one can fit a function $E(\text{var}_{\bth}[\h]) = E_0 + b\text{var}_{\bth}[\h]$ due to the approximately linear relation between $E$ and var$_{\bth}[\h]$ in the numerical results, where $E_0 = -1.9959$ Ha in the zero-variance limit is also very accurate with an error 0.1 kcal/mol.

The quantum Lanczos results, which are conveniently calculated along with AVQITE at no additional quantum resource cost, are plotted as orange squares in Figure~\ref{fig: h4}(a) for the total energy, with the error shown in the inset of panel (b). The QL calculation converges relatively faster than AVQITE, since the reduced Lanczos Hamiltonian encodes information beyond the latest quantum state. Because the QL method is generally susceptible to numerical inaccuracies and the way to evaluate the Hamiltonian and overlap matrix in the Krylov subspace used in the QL method accumulates sizable errors, excited states cannot be directly accessed in the current QL calculations. 

To demonstrate the general applicability of the AVQITE approach in generating compact ground state ans\"atze, we perform AVQITE calculations for H$_4$ chains at bond length $R_\text{HH}=0.8$\AA, $1.5$\AA, $2.4$\AA, H$_2$O at $R_\text{OH}=0.8$\AA, $1.5$\AA, $2.4$\AA, and BeH$_2$ at $R_\text{BeH}=0.8$\AA, $1.4$\AA, and $3.6$\AA, as shown in Figure~\ref{fig: all}. This benchmark set covers a variety of directional covalent bonding and electron correlation effects, with atomic states of spin singlet, doublet and triplet in the dissociation limit. In Figure~\ref{fig: all}(a) we show the AVQITE energies of H$_4$ at three bond lengths colored in red, green and blue, which agree with the FCI results within chemical accuracy. The detailed error convergence, $E-E_\text{Exact}$, as a function of imaginary time step $\tau=N\Delta\tau$ is shown in panel (d). With increasing bond length or correlation energy, the critical imaginary time $\tau_\text{c}$, which is the time where chemical accuracy is reached, generally increases. Here the correlation energy is defined as the difference between the HF and FCI energies, which corresponds to the initial AVQITE energy error as the AVQITE ansatz starts with the HF state.  At $R_\text{HH}=2.4$\AA\, proximate to the atomic limit, the critical time step increases significantly to $\tau_\text{c}\approx 110$. However, as observed earlier in numerical step-merged QITE calculations~\cite{smqite}, one can choose a bigger step size $\Delta\tau$ for molecules at larger bond length to reduce the number of imaginary time steps $N$. In this case, $\Delta\tau$ is increased from 0.1 to 0.5 to speed up the convergence, although sizable energy fluctuations are present in the initial time steps. 

The number of controlled-NOT gates (CNOTs) generally increases in the initial steps and levels off for $\tau$ above 3, as shown in panel (g). The maximal number of CNOTs reaches 250 for H$_4$ at $R_\text{HH}=2.4$\AA. Here the number of CNOTs is estimated by the rule that each multi-qubit rotation gate $e^{-i\theta\hat{\sigma}}$ with Pauli string $\hat{\sigma}$ of length $p$ needs $2(p-1)$ CNOTs (assuming all-to-all connectivity)~\cite{nielsen2002quantum}.

Moving on to the AVQITE calculations of H$_2$O and BeH$_2$ at representative bond lengths, the error convergence behavior generally remains similar to the H$_4$ calculations, with final energies within chemical accuracy. Likewise, the number of CNOTs in the AVQITE state preparation circuit generally increases initially and levels off for $\tau >3$. The total number of CNOTs remains close to or under 400. Remarkably, the positive correlation between the number of CNOTs in the AVQITE ansatz and correlation energy, which seems to exist in the calculations of H$_4$, does not apply to H$_2$O and BeH$_2$. The AVQITE state preparation circuits in the test-set are generally over one order of magnitude shallower than the original UCCSD ansatz. In panels (g-i), we show that the number of variational parameters $N_{\bth}$ of the AVQITE ansatz is generally quite close to that of the qubit-ADAPT-VQE approach. The H$_4$ chain at $R_\text{HH}=2.4$~\AA\,marked by a blue cross symbol in Fig.~\ref{fig: all}(j) represents an interesting example, where the qubit-ADAPT-VQE becomes trapped in a local minimum with energy 15 mHa higher than the ground-state energy. The qubit-ADAPT-VQE method is implemented following references~\cite{vqe_adaptive, vqe_qubit_adaptive}. The Broyden–Fletcher–Goldfarb–Shannon (BFGS) algorithm is used to optimize the variational ansatz in a new generation with an additional parameterized unitary at the optimal solution of the previous generation. This implies that the energy cost function is optimized along a continuous parameter path, which could lead to a local minimum with vanishing gradients in a generally nonconvex high-dimensional energy landscape, as exemplified here. This is related with the general barren plateau problem where the optimization of random circuits is trapped in a flat cost function region in the parameter space~\cite{mcclean2018barren}. Various techniques to address the barren plateau problem have been proposed, including the block-identity initialization strategy~\cite{grant2019initialization}, and entanglement devised mitigation techniques~\cite{patti2020entanglement}. In fact, a modified qubit-ADAPT-VQE calculation, which successively adds unitaries in accordance with the AVQITE calculation and optimizes the ansatz initialized at the AVQITE parameter values already reaches the accuracy of 0.1mHa with the first 31 unitaries out the total 33 unitaries in the AVQITE ansatz. This implies a proper combination of qubit-ADAPT-VQE and AVQITE can be mutually beneficial.

\begin{figure*}[h]
	\centering
	\includegraphics[width=0.5\linewidth]{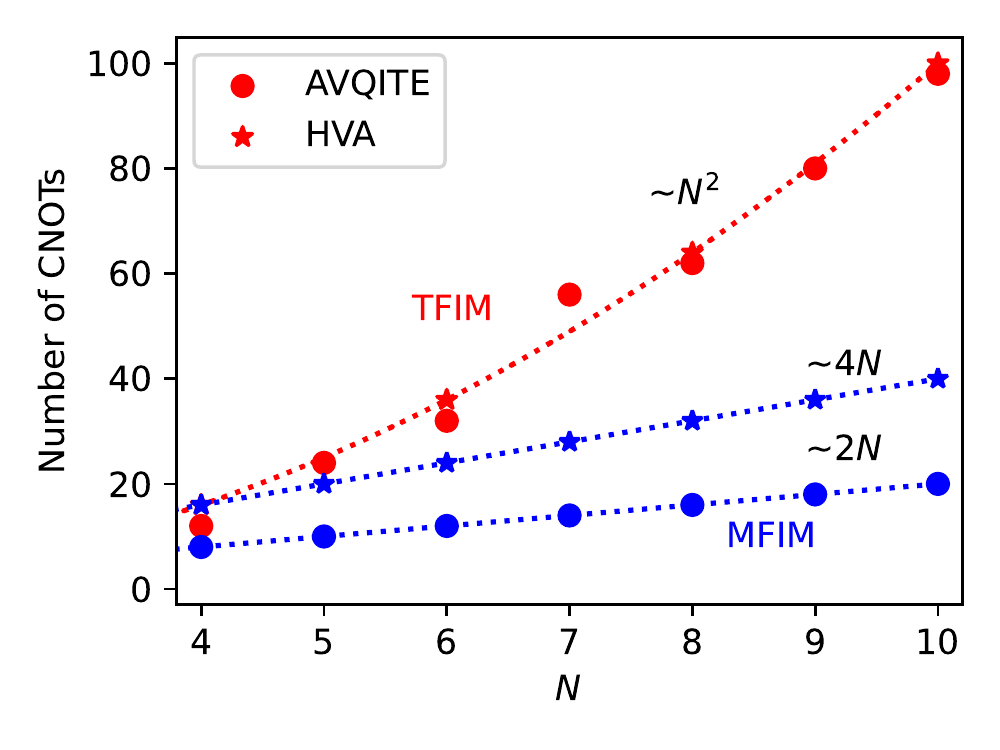}
	\caption{
	\textbf{Number of CNOT gates ($N_\text{cx}$) in the AVQITE circuit for ground state preparation of local spin model as a function of model size $N$.} AVQITE circuit prepares the quantum critical state of TFIM with $N_\text{cx}$ scaling quadratically with $N$, similar to HVA. Setting finite $h_z=0.5$ in the MFIM, the scaling reduces to linear, with a prefactor of 4 for HVA and 2 for AVQITE. The counting of CNOTs assumes cyclic connectivity of qubits as the layout of spin chain model in periodic boundary condition.
	}
	\label{fig: mfim}
\end{figure*}

\section{System-size scaling of AVQITE circuit complexity}
\label{sec:SpinModels}
Quantum resources for AVQITE calculations have been analysed in Sec.~\ref{sec: implementation}. A measure of NISQ circuit complexity is the number of CNOTs $N_\text{cx}$, which is related to the number of parameters $N_{\bth}$ associated with multi-qubit Pauli rotation gates. It is assumed that $N_{\bth}$ grows polynomially as $N^q$ with system size $N$, where the exact order $q$ is tied to ground state complexity of the system. Because of the limitations of classical simulations, it is difficult to show directly the system-size scaling of $N_\text{cx}$ in quantum chemistry calculations. Instead, we apply AVQITE to prepare ground states of local spin chain models, where the scaling of $N_\text{cx}$ is more easily accessible. We find that $N_\text{cx}$ AVQITE scales quadratically or linearly, depending on how close the system is to a quantum critical point. This scaling is similar to VQE using the well-established Hamiltonian variational ansatz (HVA) for local spin models~\cite{wecker2015_trotterizedsp, ho2019efficient, wiersema2020exploring}. We thus conjecture that a favorable polynomial order of system-size scaling for the adaptive preparation of ground states carries over to complex quantum chemistry problems, which should be addressed in future work. 

To establish the scaling of $N_{\text{cx}}$ with system size, we consider the mixed-field Ising model (MFIM):
\be
   \h = -J\sum^{N}_{i=1} \hat{Z}_{i-1} \hat{Z}_i - \sum^{N}_{i=1} \left( h_x\, \hat{X}_i + h_z\, \hat{Z}_i\right),
\ee
with $Z_{0} \equiv Z_{N}$ for periodic boundary conditions. Energy is measured in units where $J=1$. When setting the longitudinal field $h_z = 0$, the MFIM reduces to the integrable transverse-field Ising model (TFIM). The competition between ferromagnetic and paramagnetic phases is controlled by the transverse field $h_x$ of the model. When $h_x = 1$, the TFIM is at a quantum critical point. We apply AVQITE to prepare the ground state of the TFIM at the quantum critical point $(h_x, h_z) = (1, 0)$, and of the MFIM at $(h_x, h_z) = (1, 0.5)$, with $4\le N \le 10$. The following complete operator pool is adopted for the calculation~\cite{MayhallQubitAVQE, zhang2021adaptive}:
\be
\mathscr{P} = \{Y_i\}^{N}_{i=1}\cup\{Y_{i-1}Z_{i}\}^{N}_{i=1}\cup\{Z_{i-1}Y_{i}\}^{N}_{i=1},
\ee
which is composed of only local one-qubit and two-qubit Pauli strings with a single $Y$ operator. Figure~\ref{fig: mfim} shows the system-size dependence of $N_\text{cx}$ of the AVQITE circuit which prepares the ground state with fidelity $f > 99.9\%$. $N_\text{cx}$ grows as $N^2$ for preparation of the quantum critical state of the TFIM. The finite longitudinal field $h_z=0.5$ in MFIM breaks the integrability of the TFIM. However, it also drives the system away from the quantum critical point. Consequently, the ground state of the MFIM becomes less entangled, and can be prepared by an AVQITE circuit with $N_\text{cx} = 2N$. The above circuit analysis for $N_\text{cx}$ assumes that the qubit connectivity is cyclic, like the spin-site configuration of the model, such that each two-qubit rotation gate requires 2 CNOT gates.

The favorable scaling of $N_\text{cx}$ in preparing the ground state of local spin models can also be obtained using HVA, as shown in Fig.~\ref{fig: mfim}. Perfect fidelity is obtained with an $N/2$-layer ansatz for a TFIM with even system size $N$~\cite{ho2019efficient, wiersema2020exploring}:
\be
\ket{\Psi[\bth]} = \prod_{r=1}^{N/2}e^{-i\theta_{r2}\h_2}e^{-i\theta_{r1}\h_1}\ket{\Psi_0},
\ee
with $\h_1 \equiv -\sum^{N}_{i=1} \hat{Z}_{i-1} \hat{Z}_i$ and $\h_2 \equiv -\sum^{N}_{i=1} \hat{X}_i$. The reference state is chosen to be the ground state of $\h_1$: $\ket{\Psi_0} = \otimes_{i=1}^{N}\ket{+}$, with $\ket{+}=(\ket{\up}+\ket{\dw})/\sqrt{2}$. The CNOT gates in the $N/2$-layer HVA circuit can be simply counted as $N_\text{cx}= 2N \times N/2=N^2$. For the MFIM, we find that a constant two-layer HVA ansatz of the following form can prepare the ground state with similar fidelity $f > 99.9\%$:
\be
\ket{\Psi[\bth]} = \prod_{r=1}^{2}e^{-i\theta_{r3}\h_3}e^{-i\theta_{r2}\h_2}e^{-i\theta_{r1}\h_1}\ket{\Psi_0}.
\ee
Here we define $\h_3 \equiv -\sum^{N}_{i=1} \hat{Z}_i$, and choose the ground state of $\h_3$, $\ket{\Psi_0} = \otimes_{i=1}^{N}\ket{\up}$, as the reference state. The number of CNOT gates can be similarly estimated as $N_\text{cx}= 2N \times 2=4N$.

For both local spin models considered, the ground state preparation circuits for AVQITE and HVA have similar complexity as measured by $N_\text{cx}$. Nevertheless, they approach the ground state along different paths. AVQITE follows the QITE path starting with the reference state, and the imaginary time evolution approach is guaranteed to converge to the ground state~\cite{qite_chan20}. Compared with the VQE-HVA approach, the operator pool screening procedure to adaptively expand the ansatz introduces substantial overhead for AVQITE calculations. In contrast, proper parameter initialization is important for VQE-HVA calculations. In fact, VQE-HVA calculations must not start from state $\ket{\Psi_0}$, as otherwise the gradient vanishes. This can be easily achieved by initializing the parameters to nonzero values. For random parameter initialization, a finite fraction of VQE calculations on the TFIM with the $N/2$-layer HVA ansatz fail to converge~\cite{wiersema2020exploring}.

\section{Conclusion}
The adaptive variational quantum imaginary time evolution (AVQITE) approach is developed by generalizing the recently proposed adaptive quantum dynamics simulation (AVQDS) method~\cite{AVQDS} from real time to imaginary time. We present benchmark quantum chemistry calculations on H$_4$, H$_2$O and BeH$_2$ with different chemical bonding character and atomic spin multiplicity upon dissociation limit. They demonstrate the general applicability of AVQITE to finding accurate and compact variational ans\"atze for interacting many-electron models. The key advantage of AVQITE over adaptive VQE approaches is that it bypasses the complicated nonconvex optimization problem in high-dimensional parameter space by performing energy cost function minimization along a single imaginary time axis. AVQITE can also be applied to reduced ``low energy'' models in quantum chemistry such as those that arise within the complete active space self-consistent field method (CASSCF)~\cite{cramer2013essentials}. More generally, it can be used as an impurity solver for quantum embedding approaches like the rotationally invariant Gutzwiller embedding and density-matrix embedding methods~\cite{gqce, ga_pu, ga_uo2, knizia2012dmet, knizia2013dmet, sun2016quantum, ga_dmet}, extending its applicability to larger molecules and solid state materials. With the favorable polynomial scaling of the AVQITE calculations which automatically generates compact variational ans\"atze of high accuracy, we envision feasible follow-up applications of this method on NISQ devices to calculate ground-state properties of spin models and molecules. Excited states are accessible if AVQITE is used with cost functions related to the energy variance, such as in the folded spectrum method~\cite{vqe, vqe_theory, zhang2021adaptive, macdonald1934modified, wang1994solving}.



\medskip
\textbf{Acknowledgements} \par
This work was supported by the U.S. Department of Energy (DOE), Office of Science, Basic Energy Sciences, Materials Science and Engineering Division. The research was performed at the Ames Laboratory, which is operated for the U.S. DOE by Iowa State University under Contract No. DE-AC02-07CH11358.

\medskip

%
\bibliographystyle{MSP}
\bibliography{refabbrev, ref}








\end{document}